\documentclass[iicol, sn-basic]{sn-jnl}

\usepackage{graphicx}

\usepackage{amsmath}
\usepackage{amsfonts}
\usepackage{amssymb}
\usepackage{dsfont}

\usepackage{graphicx}
\usepackage{bm}

\usepackage{color}
\usepackage{xcolor}
\usepackage{siunitx}

\newcommand{\Lfunc}{\mathcal{L}}
\newcommand{\m}[1]{\ensuremath{\mathrm{#1}}}

\newcommand{\ket}[1]{\vert #1\rangle} 
\newcommand{\braket}[1]{\langle #1\rangle} 
\newcommand{\bra}[1]{\langle #1\vert }

\begin{document}

\title{Data re-uploading with a single qudit}

\author[1]{\fnm{Noah L.} \sur{Wach}} 
\author[2,3]{\fnm{Manuel S.} \sur{Rudolph}}
\author[1,4]{\fnm{Fred} \sur{Jendrzejewski}}
\author*[5]{\fnm{Sebastian} \sur{Schmitt}}\email{sebastian.schmitt@honda-ri.de}

\affil[1]{\orgdiv{Kirchhoff-Institut  f\"ur  Physik}, \orgname{Universit\"at  Heidelberg}, \orgaddress{\street{Im  Neuenheimer  Feld  227}, \city{Heidelberg}, \postcode{69120}, \country{Germany}}}

\affil[2]{\orgname{Zapata Computing Canada Inc.}, \orgaddress{\street{25 Adelaide St E}, \city{Toronto}, \postcode{M5C3A1},  \country{Canada}}}
\affil[3]{\orgdiv{Institute of Physics}, \orgname{Ecole Polytechnique Fédérale de Lausanne (EPFL)}, \orgaddress{\street{Station 3}, \city{Lausanne}, \postcode{1015},  \country{Switzerland}}}
\affil[4]{\orgname{Alqor UG (haftungsbeschr\"ankt)}, \orgaddress{\street{Alexanderstr. 65}, \city{Frankfurt am Main}, \postcode{60489}, \country{Germany}}}
\affil[5]{\orgname{Honda Research Institute Europe GmbH}, \orgaddress{\street{Carl-Legien-Str.\ 30}, \city{Offenbach}, \postcode{63073},  \country{Germany}}}

\abstract{
Quantum two-level systems, i.e. qubits, form the basis for most quantum machine learning approaches that have been proposed throughout the years. However, higher dimensional quantum systems constitute a promising alternative and are increasingly explored in theory and practice.
Here, we explore the capabilities of multi-level quantum systems, so-called qudits, for their use in a quantum machine learning context. 
We formulate classification and regression problems with the data re-uploading approach and demonstrate that a quantum circuit operating on a single qudit is able to successfully learn highly non-linear decision boundaries of classification problems such as the MNIST digit recognition problem. 
We demonstrate that the performance strongly depends on the relation between the qudit states representing the labels and the structure of labels in the training data set. 
Such a bias can lead to substantial performance improvement over qubit-based circuits in cases where the labels, the qudit states and the operators employed to encode the data are well-aligned.
Furthermore, we elucidate the influence of the choice of the elementary operators and show that a squeezing operator is necessary to achieve good performances.
We also show that there exists a trade-off for qudit systems between the number of circuit-generating operators in each processing layer and the total number of layers needed to achieve a given accuracy.
Finally, we compare classification results from numerically exact simulations and their equivalent implementation on actual IBM quantum hardware. 
The findings of our work support the notion that qudit-based algorithms exhibit attractive traits and constitute a promising route to increasing the computational capabilities of quantum machine learning approaches.
}


\maketitle 

\section{Introduction}
In recent years, the field of
quantum machine learning has attracted much attention. There, quantum circuits are employed as central processing units for data-driven applications~\citep{biamonte_quantum_2017,schuld_supervised_2018,dunjko_machine_2018}.
While it is currently not clear whether or not quantum processing can provide a benefit on practical machine learning problems~\citep{schuld_is_2022,schuld_supervised_2021}, there has been some evidence that quantum machine learning models can outperform classical models in certain tasks~\citep{liu_rigorous_2021,sweke_quantum_2021,gyurik_establishing_2022, gyurik_towards_2022}. 
While most studies, theoretical~\citep{bharti_noisy_2022,montanaro_quantum_2016} as well as experimental~\citep{graham_multi-qubit_2022, pino_demonstration_2021, kjaergaard_superconducting_2020}, focus on quantum systems consisting of two-level quantum systems, i.e., quantum bits (qubits), quantum computing hardware and algorithms can also be based on $d$-level systems, which are typically 
called qudits~\citep{wang_qudits_2020,ringbauer_universal_2022}. Such qudit systems were shown to have advantages in specific contexts~\citep{cozzolino_highdimensional_2019,sheridan_security_2010} and have already been applied to several tasks~\citep{bravyi_hybrid_2022,deller_quantum_2023, weggemans_solving_2022}. However, a full evaluation on application-relevant tasks is still lacking.

Currently, it is not clear whether there is a fundamental advantage or disadvantage of utilizing qudit systems over their qubit counterparts for quantum machine learning or other application domains such as quantum optimization. However, qudits provide a complementary route to increasing the Hilbert space size in pursuit of better computation performance. 
This is an alternative direction to qubit systems, where the route to larger Hilbert spaces is to increase the number of qubits. 
Among other challenges, increasing the number of qubits comes with increased difficulty of engineering high-fidelity interactions between separate qubits, which are technologically challenging and typically involve larger gate errors than single qubit operations, see for example, \cite{kjaergaard_superconducting_2020}, \cite{fedorov_quantum_2022} or \cite{resch_quantum_2019}. 
In that regard, expanding the local Hilbert space dimension by utilization of qudit systems seems promising, since certain interactions between spatially separated qubits could instead be implemented by local qudit operations, and thus might prove to be more 
efficient~\citep{fischer_towards_2022,ringbauer_universal_2022}.
We discuss several of such aspects of qudit system at various parts in this work.  

We explore the usage of qudit systems in the context of the data re-uploading~\citep{perez-salinas_data_2020} quantum machine learning approach to build regression and multi-class classification models. 
In this scheme, a single qudit provides a natural way of encoding multiple classes by representing each class label as an orthogonal basis state.
We focus on the prominent data re-uploading architecture, since it was shown in the original work that a single qubit is already sufficient to implement a universal classifier.
Due to the comparably small resource requirement data re-uploading circuits were also experimentally implemented on a trapped ion device~\citep{dutta_single-qubit_2022}.    
Additionally, the encoding characteristics of these models is well understood, as it generates increasingly higher Fourier terms with more layers~\citep{schuld_effect_2021}.
Thus, it is natural to first evaluate and confirm these same properties
for a single qudit as well.
Extending these circuits to multi-qubit/multi-qudit circuits would allow for more sophisticated setups including entanglement.
However, in this work we refrain from doing this and instead focus on the most simple, non-trivial case to investigate the fundamental aspects of qubits and qudits related to basic operators and Hilbert spaces.

This paper is structured as follows: In Sec.~\ref{sec:qudits}, we introduce the mathematical description of qudits. Sec.~\ref{sec:data_re-uploading} illustrates the implementation of the data re-uploading algorithm \citep{perez-salinas_data_2020} with qudits and shows the circuit structure as well as the chosen loss functions used during training. 
Sec.~\ref{sec:experimental_setup} presents the training procedure which is done numerically and on IBMQ hardware. To be able to run our model on qubit based quantum hardware, we present a way to encode qudits with multiple qubits. In Sec.~\ref{sec:results}, we verify the expressivity of our model by testing it on a simple regression problem. We then go on and test our model on multi-class classification problems where we show an intrinsic bias between the qudit state representation and the data structure. Finally, we present numerical results of the model when being trained on the MNIST handwritten digits data set~\citep{lecun_mnist_2005}. We additionally investigate an equivalent qubit-based implementation on IBMQ hardware and the effect of entangling operations on the performance of the model.

\section{Qudits}\label{sec:qudits} 

$d$ level quantum systems, typically called qudits, are a generalization of qubits to $ d>2 $  and can serve as a basis for quantum information processing. 
The  Hilbert space is spanned by $d$ orthonormal basis vectors, denoted by $ \ket{0}, \ket{1}, ... \ket{d-1} $  and 
arbitrary qudit states can be represented by the supersposition $ \ket{\psi} = \sum_{k=0}^{d-1}c_k\ket{k} $ with 
the normalization condition  $ \sum_{k=0}^{d-1}\vert c_k\vert^2 = 1 $.

Inspired by cold atom systems \citep{kasper_universal_2022}, we interpret a $d$-level qudit as a spin with total angular momentum $\ell=\tfrac{d-1}{2}$, such that the basis state $\ket{k}$ corresponds to the spin eigenstate with angular momentum $m=\tfrac{2k-d+1}{2}$.
A natural set of operations on qudits states, which is also easily implementable in experiments, is given by the angular momentum operators $\{L_x, L_y, L_z\}$. These generate rotations around the corresponding axes and obey the canonical commutation relations of the special unitary group $SU(2)$, $[L_i,L_j]=i \epsilon_{ijk}L_k$. 
The action of the angular momentum operators on the qudit basis state are given by 
\begin{align}
\label{eq:angularmometum}
 L_x\ket{k} & =  \tfrac12\big(\gamma_{d,k+1}\ket{k+1}+\gamma_{d,k-1}\ket{k-1}\big) \\
 L_y\ket{k} & = \tfrac1{2i}\big(\gamma_{d,k+1}\ket{k+1}-\gamma_{d,k-1}\ket{k-1}\big) \\
L_z\ket{k} & = \tfrac{2k-d+1}{2}\ket{k}
\end{align}
where $k\in[0,d-1]$ and with $\gamma_{d,k}=\sqrt{(d-k-1)(k+1)}$.

There are various  ways to define a universal gate set to realize arbitrary actions in the qudit Hilbert space, see e.g.\ \cite{wang_qudits_2020} and \cite{luo_geometry_2014}. 
For a single qudit, as it is considered in this work, we choose the two angular momentum operators $L_x$ and $L_z$ and the squeezing  or one-axis twisting operator $L_{z^2}=L_z^2$.
For $d>2$ this additional gate is needed to be able to  generate any state by (possibly many) repeated finite rotations as detailed in \cite{kasper_universal_2022} and \cite{giorda_universal_2003}. 
The reason for this is that the iterated commutators of these three operators generate all $d^2-1$ Hermitian basis operators, which are necessary to generate all unitary operations of the $SU(d)$ group.

Equivalent to qubits, the gates for qudit circuits are then generated by exponentiation of the basic operators which implements the rotations 
\begin{align}
    R_j(\theta) = e^{-i \theta L_j} ,
\end{align} 
with $j \in \{x, z, z^2\}$ and where $\theta\in\mathbb R$ are free parameters.

\begin{figure*}
\centering
   a\hspace*{0.3\linewidth}b\hspace*{0.5\linewidth} \\[-9mm]
   \includegraphics[width=0.24\linewidth]{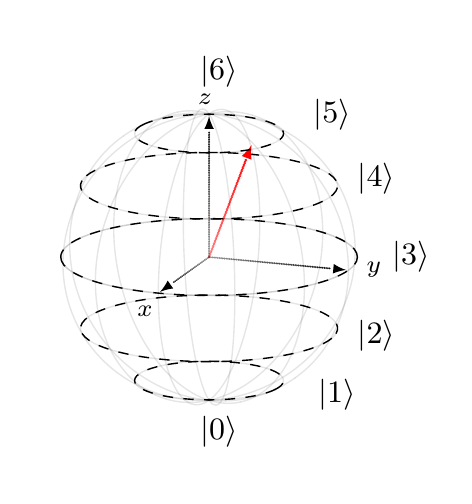} \includegraphics[width=0.5\linewidth]{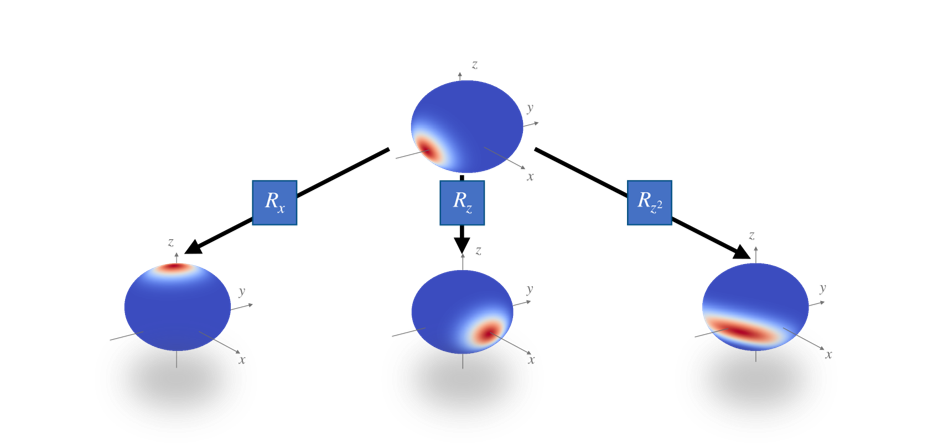}\\
   c\hspace*{0.4\linewidth}\hspace*{0.41\linewidth} \\[-6mm]
     \includegraphics[width=0.9\linewidth]{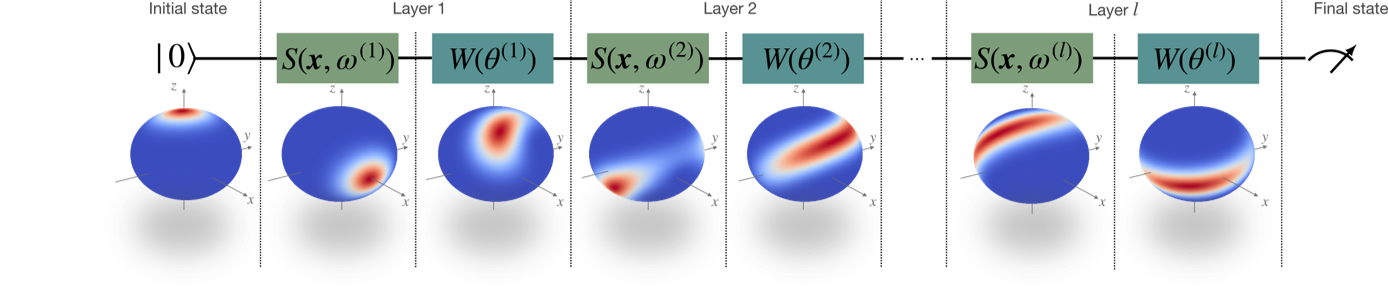}
\caption{a: Schematic representation of $d=7$ qudit states on a generalized Bloch sphere. b: Schematic illustration of the action of the three operators $R_x$ (left), $R_z$ (middle) and $R_{z^2}$ (right) on a qudits state. c: Illustration of the data re-uploading circuit structure}\label{fig:data_reuploading}
\end{figure*}
In Fig.~\ref{fig:data_reuploading} we illustrate the qudit states (panel A)  and the action of the elementary operators by showing the Husimi-Q quasi-probability distribution \citep{husimi_formal_1940} on a generalized  Bloch sphere in panel B. The pure rotation does not deform the probability distributions, but applying the squeezing operation leads to a deformation of the state. 

By choosing the operator set $\{L_x, L_z, L_z^2\}$ we  impose a ladder structure on the qudit target states. This primarily arises from the operator $L_x$, which couples each state $\ket{k}$ to its adjacent states $\ket{k+1}$ and $\ket{k-1}$, as it is illustrated in Eq.~\ref{eq:angularmometum}.    

\section{
Data re-uploading with a single qudit
}\label{sec:data_re-uploading}

We utilize a single $d$-level qudit and implement the quantum machine learning model as a data re-uploading quantum circuit \citep{perez-salinas_data_2020,jerbi_quantum_2023}.
The quantum circuit in its general form is build up from $L$ layers and encodes a quantum state, which depends on the input data $\mathbf{x}$ as
\begin{align}\label{eq:state_unitary}
    \ket{\mathbf{x},\boldsymbol{\omega},\boldsymbol{\theta}}  & 
=\prod_{l=1}^LU(\mathbf{x}, \boldsymbol{\omega}^{(l)},\boldsymbol{\theta}^{(l)}) \ket{0}\,,
\end{align}
where $\ket0$ is the initial state. The characteristic of the data re-uploading architecture is that the unitary operation of each layer $l$ encompasses data dependent unitaries, which are parametrized by the scaling parameters $\boldsymbol{\omega}^{(l)}$ and data independent operations with free parameters $\boldsymbol{\theta}^{(l)}$.

We tested several layer structures and found that they all produce similar results. In the following we report results for two architectures.  The first one is inspired by classical Euler rotations, where the unitaries of each layer have the structure:
\begin{align}
\label{eq:DRUL_euler1}
    U(\mathbf{x}, \boldsymbol{\omega}^{(l)},\boldsymbol{\theta}^{(l)}) = W(\boldsymbol{\theta}^{(l)}) S(\mathbf{x}, \mathbf{\omega}^{(l)}) \,.
\end{align}
The data encoding block $S$ of each layer consists of alternating $x$ and $z$ rotations,
\begin{align}
\label{eq:DRUL_euler2}
    S(\mathbf{x}, \bm\omega^{(l)}) = R_\alpha(x_D\omega_D^{(l)} )\cdots    {R}_z(x_2\omega_2^{(l)}){R}_x(x_1\omega_1^{(l)})
\end{align}
where the number of rotations is determined by the dimensionality $D$ of the input data and consequently $\alpha=x$ ($\alpha=z$) in case $D$ is odd (even). 
The alternating $x$ and $z$ rotations is chosen to ensure that these are non-commuting. This is necessary to distinguish between the individual dimensions  of the data vector. If only $x$ (or $z$) rotations were chosen to encode the data, the classifier would only be able to learn the sum of the input data.

The data-independent block in each layer is composed of a sequence of three rotations followed by a squeezing gate $R_{z^2}$ as the last operation, i.e.\
\begin{align}\label{eq:DRUL_euler3}
     W(\boldsymbol{\theta}^{(l)}) = {R}_{z^2}(\theta^{(l)}_4){R}_{x}(\theta^{(l)}_3){R}_{z}(\theta^{(l)}_2){R}_{x}(\theta^{(l)}_1)
\end{align}
 The first three operations give rise to Euler angles, starting from $\ket{0}$, and thus the freedom to create overlap with an arbitrary qudit state. 
The total number of adjustable parameters of a $L$-layer model  is  given by $(4+D) L$. 
The action of this circuit structure is illustrated in Fig.~\ref{fig:data_reuploading} C. 

The second architecture considered here is inspired by the simplified form presented by \cite{perez-salinas_data_2020} and has the structure:
\begin{align}
    \label{eq:simleCirc}
    U(\mathbf{x}, \boldsymbol{\omega}^{(l)},\boldsymbol{\theta}^{(l)}) = e^{-i\sum_{j}^D (\theta_j^{(l)}+\omega_j^{(l)} x_j) L_{c(j)} -i \theta_{D+1}^{(l)} L_{z^2} }\,.
\end{align}
Here the first term in the exponent is the sum over the three angular momentum operators, i.e.\ the generators of the $SU(2)$,  and the function $c(j)=(j\mod 3)$ selects one of them. The second term in the exponent is a generalization of the original simplified architecture by including the squeezing operator in each layer.           
The total number of adjustable parameters for an $L$-layer model of this structure  is  given by $(2D+1) L$.

We like to point out that in this form both architectures implement the aforementioned ladder structure, since the data encoding is done with the angular momentum operators, that only couple each qudit basis state to its adjacent states. 

In our work we approach supervised classification and regression tasks by using qudit-based quantum circuits. 
In a supervised learning setting, there exists a set of $N$ training samples $(\mathbf{x}_i,y_i)$ with $i=1,\dots, N$, which consist of pairs of input samples $\mathbf{x}$ with corresponding output values $y$. 
The input samples are real-valued $D$-dimensional vectors $\mathbf{x}\in\mathbb{R}^D$ while the output values are either real numbers or a finite set of integers for regression and classification tasks, respectively. 

For classification problems the output values $y\in(0, 1,\dots,d-1)$ indicate to which of the $d$ classes the data sample $\textbf{x}$ belongs. 
In the quantum formulation, each class label is represented by a basis state of the $d$-level qudit $\ket{y}$ with $y=0,\dots,d-1$. The model prediction, i.e., the probability that a data sample $\mathbf{x}$ belongs to class $y$, can then conveniently be calculated from the overlap of the label state and the qudit wave function obtained from the  quantum circuit with input $\mathbf{x}$, 
\begin{align}
\label{eq:overlap}
    P(y\vert \mathbf{x},\boldsymbol{\omega},\boldsymbol{\theta}) = \vert \braket{ y\vert \psi(\mathbf{x},\boldsymbol{\omega},\boldsymbol{\theta}) }\vert ^2\,.
\end{align}
Training the quantum circuit is achieved by minimizing a loss function over the given training data set $\mathcal{D} = \{ (\textbf{x}_i,y_i)\}_{i=1,\dots, N}$.
The overlap of Eq.~\eqref{eq:overlap} is the basis for formulating the  mean squared error (MSE) loss function,
\begin{align}\label{eq:mse}
    \Lfunc_{\text{MSE}}(\boldsymbol{\omega},\boldsymbol{\theta})  &=\frac{1}{N}\sum_{i=1}^N \big(\braket{ \bar{y}_i}- y_i\big)^2
 \end{align}
 with the average predicted label of the quantum model
\begin{align}\label{eq:averageoutput}
\braket{ \bar{y}_i} = \sum_{y=0}^{d-1} y \,P(y \vert \mathbf{x}_i,\boldsymbol{\omega},\boldsymbol{\theta})\,.
 \end{align}
 Another popular choice is the overlap loss as used in \cite{perez-salinas_data_2020},
\begin{align}
    \Lfunc_\text{overlap}(\boldsymbol{\omega},\boldsymbol{\theta} )  = 
      \sum_{i=1}^N \big(1- P(y_i \vert  \mathbf{x}_i,\boldsymbol{\omega},\boldsymbol{\theta} )\big).
\end{align}
The learning procedure amounts to adjusting the parameters of the quantum circuit $(\boldsymbol{\omega},\boldsymbol{\theta})$ in order to minimize the loss function, which is done by running a classical optimization algorithm.

After the quantum circuit has been trained its accuracy is evaluated by analyzing its predictions of the output variables on a test data set, i.e.\ a set of  data samples not used during the training procedure. 
For a classification task the predicted output labels is given by the basis state with the highest probability in the quantum state with the corresponding input data sample, i.e.
\begin{align}
\label{eq:predictedclassification}
    y_i^\m{predicted}=\m{argmax}_yP(y\vert \mathbf{x}_i,\boldsymbol{\omega}_\m{opt},\boldsymbol{\theta}_\m{opt})\,,
\end{align}
where  $(\boldsymbol{\omega}_\m{opt},\boldsymbol{\theta}_\m{opt})$ are the optimized values of the circuit parameters.
The accuracy of the trained model can then be evaluated by calculating the fraction of correctly predicted labels in the test data set,
 \begin{align}
    \text{Accuracy} = \frac1{N_\m{test}}\sum_{i\in\mathcal{D}_\text{test}} \delta_{y_i^\m{predicted},y_i} 
    \,,
\end{align}
where  $\mathcal{D}_\text{test}$ denotes the test data set which contains $N_\m{test}$ data samples and $\delta_{a,b}$ is the Kronecker delta.

The MSE loss of Eq.~\eqref{eq:mse} is also suitable to learn regression tasks. In that case the output variables are finite-range continuous variables, $y\in[0,d-1]$. The resulting qudit state of the trained quantum circuit is then a superposition of basis states and the predicted output value for input $\mathbf{x}_i$ is calculated as the expectation value of Eq.~\eqref{eq:averageoutput},
\begin{align}\
\label{eq:predictedregression}
    \braket{\bar {y}_i}^\m{predicted}=\sum_{y=0}^{d-1} y \,P( y \vert \mathbf{x}_i,\boldsymbol{\omega}_\m{opt},\boldsymbol{\theta}_\m{opt})\,.
\end{align}  

For simulations running on actual quantum hardware, the probability distribution of Eq.~\eqref{eq:overlap} is estimated by performing a  finite number of measurement shots and recording the measurement results as a histogram. This approximative distribution is then used to calculate the predicted output values of Eqs.~\eqref{eq:predictedclassification} and  \eqref{eq:predictedregression}.

\section{Experimental setup}\label{sec:experimental_setup}
We perform the training of the quantum circuit utilizing exact numerical simulations of the qudit states and quantum gates.
The parameter values of $\boldsymbol{\omega}$ and $\boldsymbol{\theta}$ are initialized randomly in the range $[-\pi, \pi]$. As classical optimization algorithms  for minimizing the loss function  we employ the ADAM optimizer~\citep{kingma_adam_2014} paired with the exact gradients calculated by the automatic differentiation package JAX~\citep{bradbury_jax_2018}. Additionally, we also used L-BFGS-B and Powell optimization approaches from the \texttt{scipy} library \citep{virtanen_scipy_2020}, where we do not utilize automatic differentiation. Apart from differences in the run time for the training, the results obtained with all approaches were always comparable. 
If not stated otherwise, all models are trained on randomly selected data sets with size $N=750$. The performance is evaluated on a separate randomly selected test data set, which contains $\tfrac N3=250$ data samples. 
In order to obtain reliable statistics of the results, we run 60 different simulations for each setting, each time randomly varying the data set and the model initialization. To visualize the decision boundaries of the trained classifiers, we utilize the visualization library \textit{orqviz}~\citep{rudolph_orqviz_2021}.

In addition to the numerically exact simulation, in \ref{sec:ibmq} we re-train and evaluate the qudit quantum circuit on the IBM \texttt{ibmq\_lima} hardware using the qiskit \citep{treinish_qiskitqiskit_2023} framework. This allows us to estimate the impact of gate errors and noise of actual NISQ hardware on the learning performance. 
As the IBM hardware naturally operates on qubits, we employ a mapping of the $d$-level qudit Hilbert space to $d-1$ qubits, which is inspired by cold atom systems~\citep{kasper_universal_2022,santra_squeezing_2022}. 
The qudit basis state $\ket{k}$ is represented by the qubit Dicke-state $\ket{D^{d-1}_k}$, i.e.\  $\ket k \to\ket{D^{d-1}_k}$ \citep{gasieniec_deterministic_2019}. The $k^{th}$ Dicke state $\ket{D^{d-1}_k}$ of $d-1$ qubits is given by the equal superposition of all states which have $k$ qubits in the state $\ket{1}$ and $d-1-k$ in state $\ket 0$, i.e.\ 
\begin{align}
     \ket{D^{d-1}_k}=\begin{pmatrix}d-1\\k\end{pmatrix}^{-\tfrac{1}{2}}\sum_{x\in\{0,1\}^{d-1}, hw(x)=k } \ket{x}\,.
\end{align}
where $hw(x)$ indicates the Hamming weight of string $x$, i.e.\ the number of 1's in $x$.
The angular momentum operators for the $d-1$ qubit states are defined as the sum over the single-qubit operators,
$L^\text{tot}_a=\sum_{j=0}^{d-1} L_a^j$ with $a=\{x,y,z\}$, and they act as described by Eqs.~\eqref{eq:angularmometum} on the Dicke states. 
In particular, the Dicke state $\ket{D^{d-1}_k}$ and the qudit state $\ket{k}$ have the same $z$-component of  the angular momentum, i.e.\ $ L^\text{tot}_z\ket{D^{d-1}_k}=\tfrac{2k-d+1}{2} \ket{D^{d-1}_k}$.

In this representation the squeezing operation  consists of all pairwise qubit interactions,
\begin{align}
L^\text{tot}_{z^2}=\big(L^\text{tot}_{z}\big)^2=\sum_{i,j} L_z^i L_z^j
\end{align}
which is implemented as two-qubit $zz$-rotation gates on the hardware~\citep{treinish_qiskitqiskit_2023}. 
From this form it is clear that the squeezing operator has the capability to generate correlation and entanglement between the individual qubits~\citep{santra_squeezing_2022}.
On the qubit device, we start from the optimized parameters found by the training with the exact simulation and re-train the model on the actual hardware. There, we perform 512 measurement shots for each qiskit circuit evaluation.

For comparison we also report the accumulated results on classification tasks of a standard classical machine learning approach, namely the \texttt{scikit-learn} \citep{pedregosa_scikit-learn_2018} implementation of the random forest (RF) classifier with 100 estimators, a $k$-nearest neighbor classifier with $k=3$ ($k$nn) and a support vector classifier (SVC). The performance of these approaches were always comparable and we report the cumulative results of running each algorithm 50 times on randomized data stets.

\section{Results}\label{sec:results}
\begin{figure}
\centering 
 \includegraphics[width=0.23\textwidth]{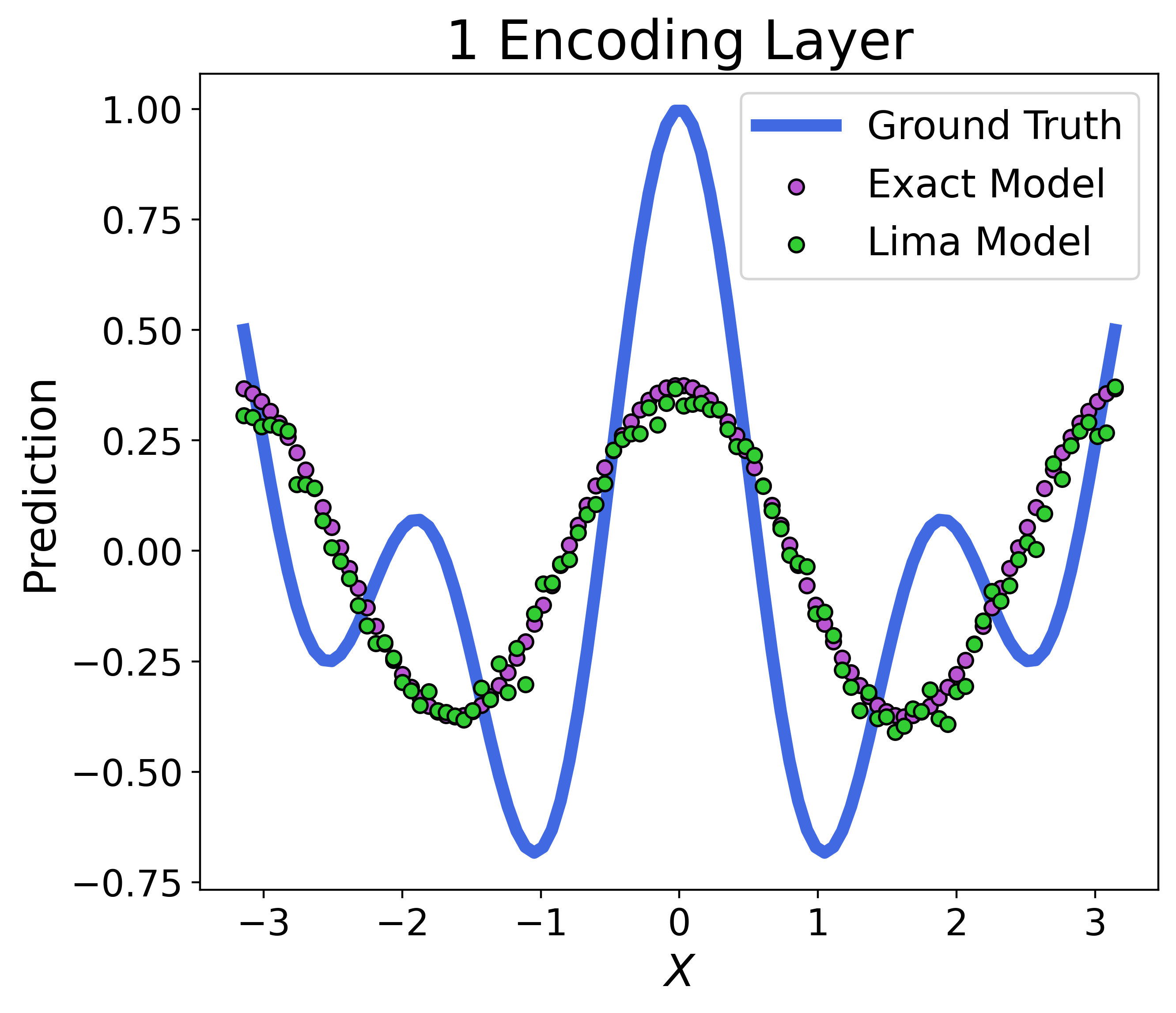}
  \includegraphics[width=0.23\textwidth]{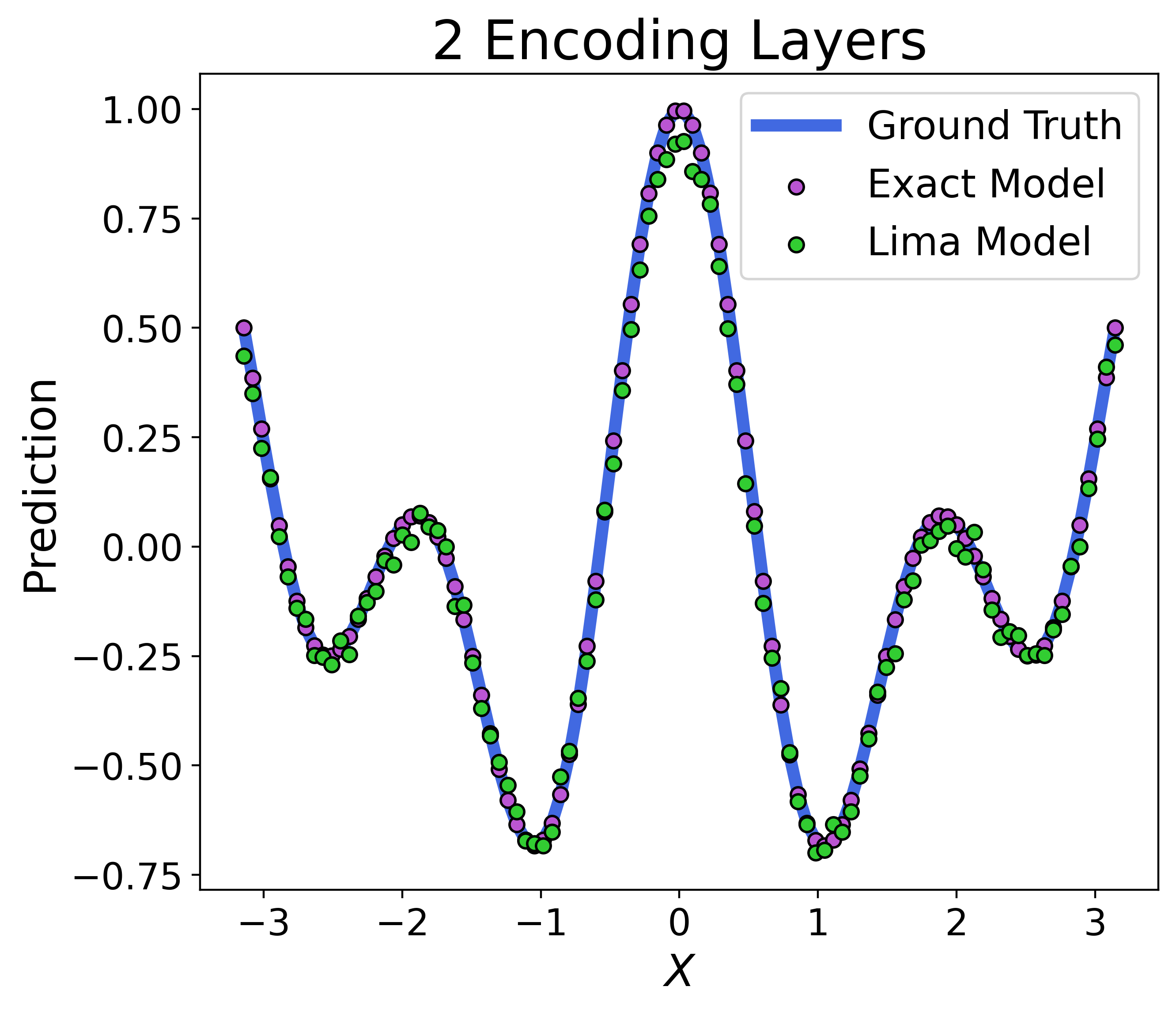}
      \caption{Results of trained data re-uploading models with structure of Eqs.~\eqref{eq:DRUL_euler1}-\eqref{eq:DRUL_euler3} using a $d=3$ qutrit with one (left) and two (right) layers for the regression task of learning the function in Eq.~\eqref{eq:simple1D} (blue line). 
      The purple dots indicate the results of the numerically exact model simulation obtained and the green dots the results from  re-training and evaluation on the IBM \texttt{ibmq\_lima} hardware
    }\label{Fig:03}
\end{figure} 

\subsection{Expressivity in one dimension}
First, we examine the expressivity of the data re-uploading circuit with a single qudit on a continuous one-dimensional regression problem  analogous to ~\cite{schuld_effect_2021}. 
We train a model with one qutrit, i.e.\ a $d=3$ qudit, to learn the simple function  
\begin{align}\label{eq:simple1D}
    f(x)=\tfrac12 \big(\cos{(2 x)} + \cos{(3.5 x)}\big)
\end{align}
with \(x \in [-\pi, \pi]\).
As the model output we use the expectation value of Eq.~\eqref{eq:averageoutput} but shift it to match the data range, i.e.\  $f_i^\m{predicted}=\braket{\bar{y_i}}-1$. For training the quantum circuit we employ the
MSE loss function of Eq.~\eqref{eq:mse} and use a training set of 100 linearly distributed samples.

We show the results of a $L=1$ layer model in the left panel of Fig.~\ref{Fig:03}. 
It is evident that this model is not able to learn the function properly. This confirms the previous insight for qubit circuits~\citep{schuld_effect_2021}, that data re-uploading models learn truncated Fourier series, and that a circuit containing only one angle parametrized by the input can only encode one Fourier component. 
In the right panel we show the results of a model with $L=2$ layers.  As expected, this model is now able to learn the function with two Fourier coefficients exactly. The derivation is analogous to the qubit case~\citep{schuld_effect_2021}, but is not shown in this work.
Apart from the results obtained by numerically exact simulations, we also include results from the evaluation on the IBM \texttt{ibmq\_lima} hardware (green dots in the plot). These predicted values are generally noisy which result from the inherent errors in the quantum circuit, as well as the shot noise from the finite number of samples used to estimate the probability distribution of Eq.~\eqref{eq:overlap}. 

\subsection{Two-dimensional classification tasks}
\begin{figure*}
    \includegraphics[width=0.95\textwidth]{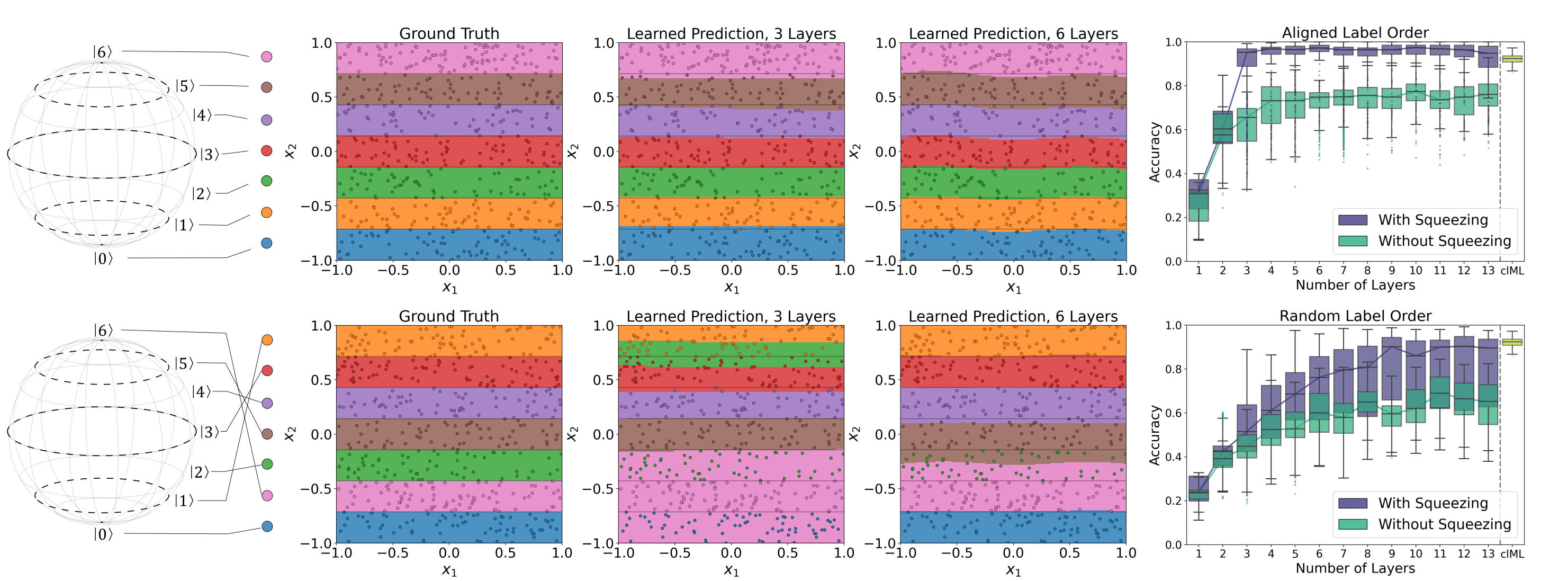}
   \caption{Results of the seven-class horizontal stripe classification problem with circuit structure of Eqs.~\eqref{eq:DRUL_euler1}-\eqref{eq:DRUL_euler3}. 
   In the upper row the qudit states are aligned with the class labels, while in the lower row the class labels are randomly assigned to the qudit states as illustrated in the pictures on the left.
   The first stripe pattern shows the  ground truth and the training data samples. 
   The middle and right stripe patterns show the classification regions obtained after training  with three and six layers, respectively. 
   The rightmost plots show the statistics of classification accuracy, obtained from 50 independent training runs, as a function of layers in the quantum circuit with (blue) and without (green) the squeezing operation in the $W$ blocks. 
   Colored boxes indicate 25\% and 75\% percentiles and horizontal whiskers the minimal and maximal values.
   In the right panels, cML indicates the results of the classical machine learning approaches
   }\label{fig:StraightLines}
\end{figure*}

Inspired by the original benchmark of the data re-uploading algorithm~\citep{perez-salinas_data_2020}, we investigate the  performance of the qudit circuit with the structure of Eqs.~\eqref{eq:DRUL_euler1}-\eqref{eq:DRUL_euler3} on various two-dimensional multi-class classification problems. 
The data samples are located on a two-dimensional square $\mathbf{x}=(x_1,x_2) \in [-1, 1]^2$, where each sample is associated with one out of $d$ classes. 
In the first problem setting, the classes are arranged in  parallel horizontal stripes. 
Fig.~\ref{fig:StraightLines} shows the results for seven classes, where we used a  $d=7$ qudit to represent the classes in the quantum circuit.

In the upper row of the figure, we show results from the case where we align the qudit states with the labels in such a way, that the $z$-components of the spin are in order with the class label. In that case, adjacent classes are represented by qudit states with adjacent $z$-spin values. Under these circumstances, the model has a strong inductive bias towards the data set caused by the chosen ladder structure.
This is illustrated in the leftmost figure where we draw the qudit states on the generalized Bloch sphere.
The lowest qudit state $\ket 0$ is associated with the class of the lowest (blue) stripe, the second lowest qudit $\ket 1$ with the second lowest stripe, and so on. 
The corresponding results show that the data re-uploading circuit can predict this data set almost perfectly with three or more layers, $L\geq3$ (blue graph in top right plot). 
Remarkably, the performance of the data re-uploading circuit is even better than the classical machine learning models shown on the far right of the right panel.
However, the learned classes and decision boundaries, as shown in the middle panels of the Figure, can still differ from the ground truth. This is due to the small size of the training data set, which necessarily leads to small random variations of the learned  decision boundaries, depending on the precise location of the training data close to the decision boundaries.

The importance of the squeezing operation $R_{z^2}$ in the $W$ operator of Eq.~\eqref{eq:DRUL_euler3} is highlighted by observing the massively degraded performance of the same circuits without this gate (green graph in the top right panel of Fig.~\ref{fig:StraightLines}). Without squeezing, the median accuracy saturates at around $0.7$ while with squeezing the median accuracy reaches $0.95$ and higher. 
This performance difference is attributed to the fact, that the squeezing operator is necessary to allow for the representation of arbitrary unitary operations in the qudit Hilbert space. 

 \begin{figure*}
  \includegraphics[width=0.99\linewidth]{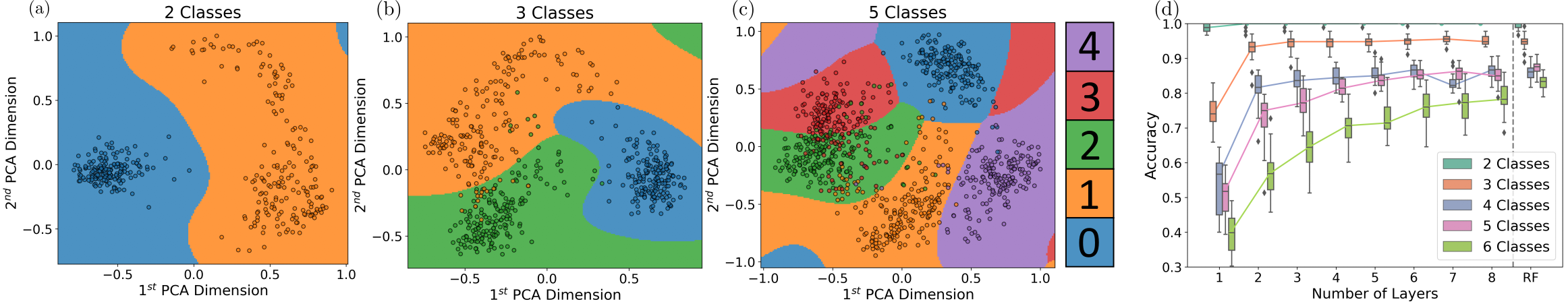}\\[-3mm]
       \caption{Learned classification regions of a single qudit classifier circuit with $L=5$ layers  trained on the first two (a), three (b) and five (c) classes (i.e.\ digits) of the MNIST data set, which was compressed down to two dimensions using PCA. 
       Each background color indicates one region associated with the same label. The training data samples are shown in the plots and colored according to their true label.
       (d) Accuracy of the qudit classifier as a function of the number of layers for various numbers of classes.  RF denotes the  random forest classifier for comparison
       }\label{fig:orqviz_mnist_pca}
\end{figure*} 

The second row of Fig.~\ref{fig:StraightLines} shows the performance when the class labels are randomly assigned to the qudit states. 
This removes the inductive bias of the model and makes the problem much harder to learn for the qudit quantum circuit, since the ladder structure used for the data encoding does not align to the data set.
Consequently, the accuracy drops significantly and many more layers are necessary to recover the performance level of the scenario with aligned labels.

The plots of the accuracies also show the accumulated results of the three classical machine learning classifiers RF, SVC and $k$nn for comparison. The performance of the quantum circuits including squeezing with randomized assignment of qudit states to labels is comparable to the classical approaches.  For aligned labels the quantum circuit even slightly outperform the classical approaches.   

We also considered several other two-dimensional multi-class classification problems and trained data re-uploading models on several horizontal stripe data sets with varying number of stripes (i.e.\ classes), as well as data sets where the stripes are not horizontal but rotated by an angle. 
We also investigated the models on data sets where the class regions are given by concentric rings with approximately  the same width and where the center of the rings is somewhere in the plane. 
The  results (not shown) on those models  were always qualitatively similar to the ones presented here. Squeezing is always necessary to achieve good performance and when the label order is aligned with the qudit states, the performance is consistently higher. 
The overall performance was slightly reduced for the tilted stripes and the concentric rings and the variance in the results was also slightly larger as compared to the horizontal stripes cases presented here.

\subsection{Classifying MNIST data} 
As the next application example, we train the model of Eqs.~\eqref{eq:DRUL_euler1}-\eqref{eq:DRUL_euler3} on subsets of a \texttt{scikit-learn}~\citep{pedregosa_scikit-learn_2018} version of the MNIST handwritten digits data set~\citep{lecun_mnist_2005}. 
This version includes down-sampled images with  $8$x$8$ pixels instead of the  $28$x$28$ pixels of the original MNIST data set. 
Since it is very computationally demanding to encode $8\times8=64$ dimensional data samples into the quantum circuit, we reduce the input dimension further using the principal component analysis (PCA)~\citep{jolliffe_principal_2016}. 

In Fig.~\ref{fig:orqviz_mnist_pca}(a)-(c) we visualize the classification boundaries of a $L=5$ layer circuit for two, three and five classes, where the input dimension is reduced to $D=2$ using a PCA. The plots show that the data re-uploading classifier is able to learn highly non-linear decision boundaries. It can also be seen that the classifier tends to produce disconnected classification regions due to the oscillatory nature of parametrized quantum circuits. The statistics of the accuracy for various classes is shown in Fig.~\ref{fig:orqviz_mnist_pca} (d) as a function of the quantum circuit layers. 
It can be observed that the accuracy,  as expected, increases with more layers, and that increasing the number of classes in the problem reduces the accuracy. For more classes, the reduction to two dimensions using PCA leads to more spacial overlap between classes, which makes the problems inherently noisy and limits the overall achievable accuracy. The overlapping classes can be directly observed in panel (c).    
Additionally, panel (d) shows the result from 50 runs of a random forest (RF) classifier. 
The accuracy of the quantum-based classifier approaches the values of the RF model with increasing number of layers.
However, the variance in the result is significantly larger for the data re-uploading circuit. One reason for this is found in the increasingly more complex classical optimization problem when increasing the number of parameters in the quantum circuit.

\subsection{Qudit vs Qubit} 

The quantum circuits we employed up to this point used a single $d$-level qudit to solve  classification problems  with $d$ classes, where each basis state encodes one class label. 
While it may appear less natural, multi-class classification problems can also be learned with a data re-uploading circuit operating with a single qubit~\citep{perez-salinas_data_2020}. 
In those approaches, the different classes are represented by single-qubit quantum states  which are chosen to be maximally orthogonal. Unless a two-class classification problems is considered, where the labels states are $\ket{0}$ and $\ket{1}$, these states cannot be fully orthogonal to each other. 
For a $d=6$ class example, we choose the eigenstates of the three spin operators, i.e.\  
$\ket{y} \in  \{\ket{0}, \ket{1}, \tfrac{1}{\sqrt{2}}(\ket{0}+\ket{1}),\tfrac{1}{\sqrt{2}}(\ket{0}-\ket{1}),\tfrac{1}{\sqrt{2}}(\ket{0}+i\ket{1}),\tfrac{1}{\sqrt{2}}(\ket{0}-i\ket{1})\}$,
as the maximally orthogonal label states.
Note that in this situation, the overlap of Eq.~\eqref{eq:overlap} as a function of the labels is no longer a proper probability distribution since $\sum_y P(y)>1$.

The natural question which arises is whether there is any difference when using qubit or qudit data re-uploading circuits for multi-class classification problems. Therefore, we study and compare the performance of a single qubit as well as a single qudit on several problems.
For the single qubit approach, we utilize the simplified data re-uploading structure of Eqs.~\eqref{eq:simleCirc}. 
Here, the squeezing operator can be removed from the circuit since it is proportional to the identity, $L_{z^2}=\tfrac{\mathds{1}}{4}$, and therefore only applies a global phase.

\begin{figure}
\includegraphics[width=0.23\textwidth]{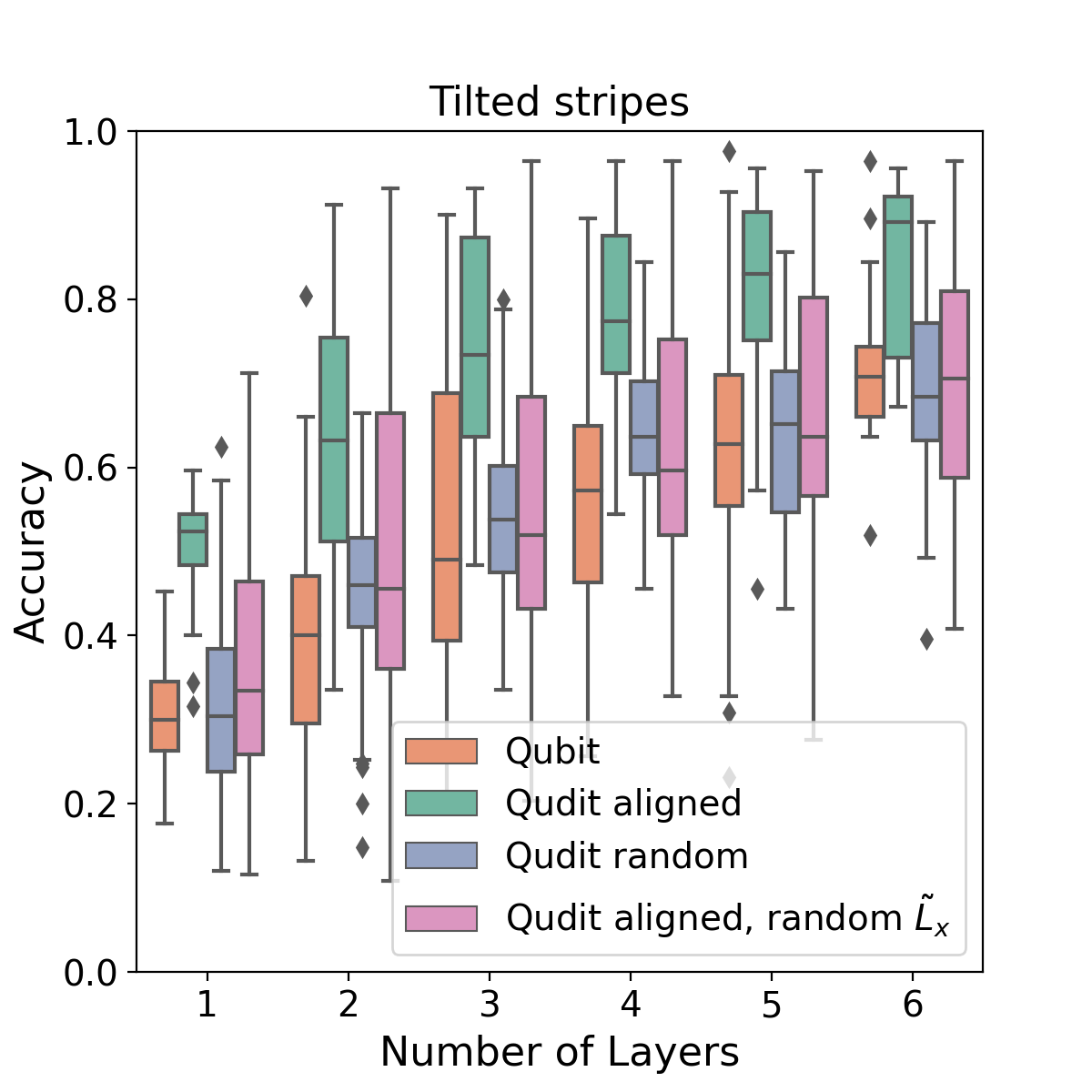}
     \includegraphics[width=0.23\textwidth]{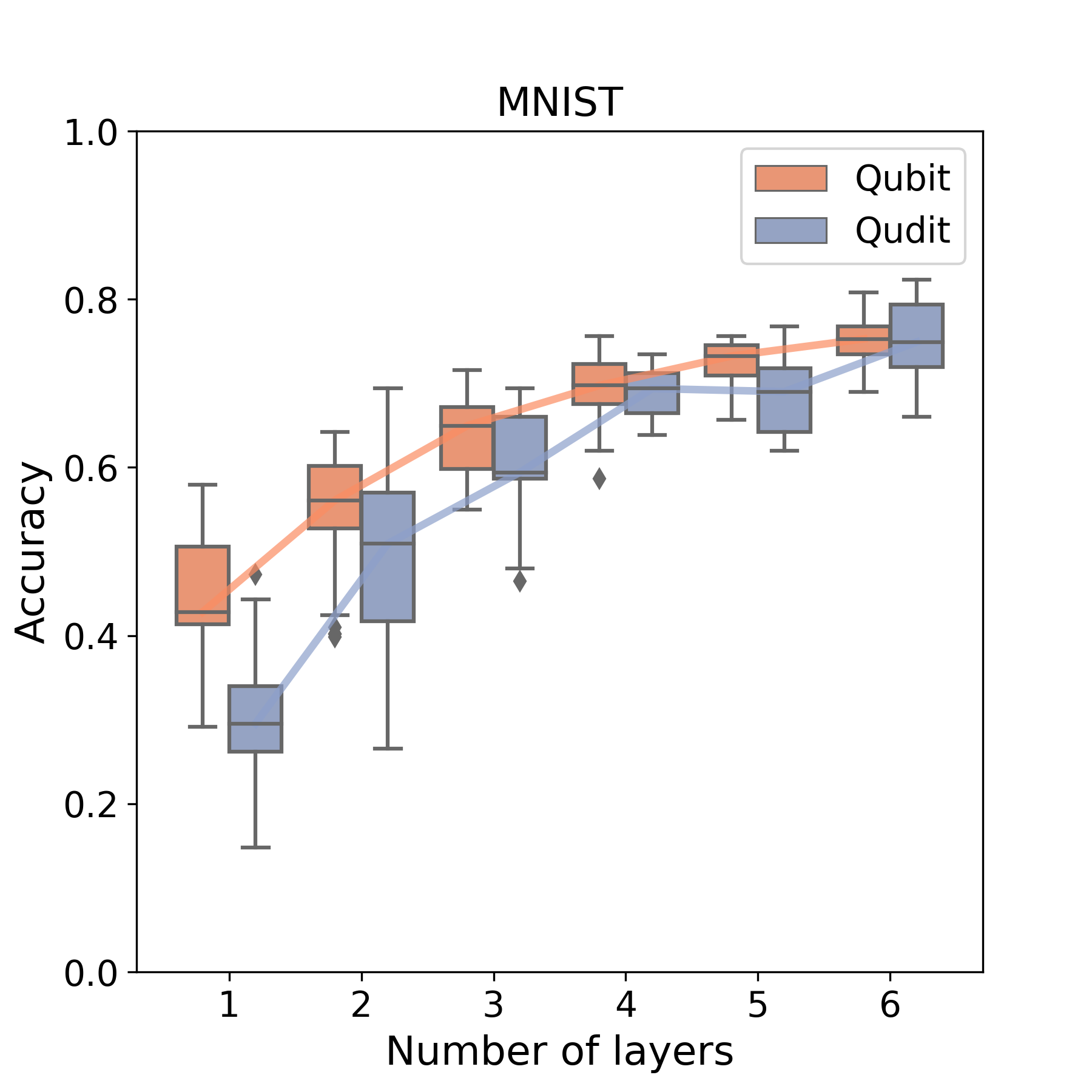}
       \caption{
       Comparison of $d=6$ multi-class classification performance of qubit and qudit data re-uploading circuits with the simplified structure of Eq.~\eqref{eq:simleCirc}.
       Left: tilted $D=2$ dimensional stripes data set with a rotation angle of $27^\circ$. 
       Qudit accuracies are shown for aligned and randomized label and state assignment, as well as with aligned labels and qudit states, but with a randomized $\tilde{L}_x$ in the encoding.
       Right: six randomly selected digits of the MNIST data set which are reduced to $D=2$ input dimensions using a PCA
       }\label{fig:qubitVSqudit}
\end{figure} 

Fig.~\ref{fig:qubitVSqudit} shows the results where the qubit and qudit models were trained on two different classification problems with $d=6$ classes and $D=2$ input dimensions. The left panel of Fig.~\ref{fig:qubitVSqudit} shows the classification accuracy as a function of the circuit layers for the tilted stripes data set, where the horizontal stripes from Fig.~\ref{fig:StraightLines} are rotated by an angle of $27^\circ$. 
It can be seen that the accuracy of the qubit and qudit circuits are comparable for the case where the class labels were randomly assigned to the qudit basis states.
However, for labels aligned with the qudit states, the performance is consistently higher for the qudit circuits. 
This result is similar to the one obtained for the horizontal stripe data set reported in Fig.~\ref{fig:StraightLines} and can be explained by the ladder structure employed in the qudit circuits. This gives rise to the inductive bias and leads to an performance improvement in case the structure of the data encoding operators are aligned with the data. 
This view is supported by a slight variation of the setup: we consider aligned labels and basis states but replace  the $L_x$ operator in the encoding scheme with a randomized version,  $\tilde{L}_x=\sum_{k}(L_{\mathcal{P}(k+1),\mathcal{P}(k)}+L_{\mathcal{P}(k-1),\mathcal{P}(k)})$.
Here, $\mathcal{P}$ denotes a random permutation of the label set $k=0,\dots,d-1$ and $L_{k,k'}$ generates a transition from state $k'$ to state $k$.
These operators do not implement a ladder structure, since the $\tilde{L}_x$ generate transitions between random states.
We indeed observe in Fig.~\ref{fig:qubitVSqudit} that the median performance then decreases to the similar level as for the situation with a ladder structure, but randomized labels.

The right panel shows the results of the qubit and qudit data re-uploading circuits for the  MNIST handwritten digits data set limited to six randomly selected digits (i.e.\ classes) and where each image is reduced  to two input dimensions using a PCA.
Here, qubit and qudit approaches are comparable in their performance, while the qubit architectures seem to slightly outperform the qudit ones, especially  for smaller circuit depths.
In this problem, the decision boundaries between the different classes are highly non-linear as discussed before. Therefore, there is no apparent favorable alignment between qudit basis states and class labels, and consequently also no inductive bias which could lead to an improved qudit performance.

In our simulations, the loss function and the classification accuracy are calculated  numerically exact with full access to the quantum state.
When running on real quantum hardware, this cannot be done and one needs to prepare and sample, i.e.\ measure, the quantum state multiple times in order to estimate the probabilities/overlaps of Eq.~\eqref{eq:overlap}. 
For the qudit system, where each class label is associated with an orthogonal basis state in the measurement basis, the statistics of the measurement outcomes directly translate to estimating the probabilities.
For qubit systems with non-orthogonal label states (i.e., for $d>2$), different measurement protocols such as quantum state tomography or quantum state discrimination~\citep{barnett_quantum_2009} need to be employed. 
For fixed number for measurement shots, these lead to a slightly increased error probability for discriminating between non-orthogonal states.   
Additionally, the difference in overlap between a true label state and  an undesired label state is less pronounced when different label states have a finite overlap. 
For qudits, this difference is always one since $P_\text{qudit}(y\vert y')=\delta_{y,y'}$ for label states $y,y'$ (here, basis states). In the qubit representation, the difference is $\leq1$ since $P_\text{qubit}(y\vert y')=\delta_{y,y'} + \sum_{y''\neq y'} c_{y''} \delta_{y,y''}$ with $c_{y}\geq 0$.
For example, in the case of $d=6$ classes considered here, $P_\text{qubit}(0\vert y')=\vert\braket{ 0\vert y' }\vert^2=\tfrac{1}{2}$ for $y'\in\{2,3,4,5\}$. 
This can lead to a reduced training signal in the loss functions using the overlap of Eq.~\eqref{eq:overlap}, since even contributions from incorrect label states have non-zero overlap and therefore reduce the loss.
Obtaining the overlaps using a finite number of measurement samples necessarily adds noise to the estimates,
which in turn makes the problem of discriminating between  correctly and wrongly predicted labels more difficult. 
As a consequence, 
the shot noise 
is expected to have a more pronounced negative effect on the training as well as prediction performance for multi-class problems when overlapping label states are used.

\subsection{Circuit structure and basic operators} 
For qubit circuits, the spin-$\frac12$ Pauli matrices  allow to represent arbitrary single-qubit unitary operations inside each layer. 
For $d$-level qudits, one instead needs to include all $d^2-1$ generators of the special unitary group in $d$ dimensions $SU(d)$, to achieve the same arbitrary control. 
However, as indicated above and shown in \cite{kasper_universal_2022} and \cite{giorda_universal_2003}, the three operators $L_x$, $L_z$ and $L_{z^2}$ are sufficient to represent any unitary operation by repeated finite rotations with multiple layers. 
This leads to the question whether there is a benefit when more than this reduced set of three operators are used in the data re-uploading circuit. We test this hypothesis by adding the following two types of operators,
\begin{align}
\label{eq:extOps}
    X_{j}&=\ket{0}\bra{0} - \ket{j}\bra{j} \qquad  (1\leq j\leq d-1) \nonumber\\
    Y_{j}&=\ket{0}\bra{j} + \ket{j}\bra{0} \, .
\end{align}
The motivation behind this choice is that these operators directly couple the initial state $\ket{0}$ to all other states $\ket{j}$ and thus may allow for a more efficient learning. 
We then use the simplified structure of Eq.~\eqref{eq:simleCirc} and replace the squeezing operator by the sum over all operators in Eq.~\eqref{eq:extOps}.  

\begin{figure}
\includegraphics[width=0.23\textwidth]{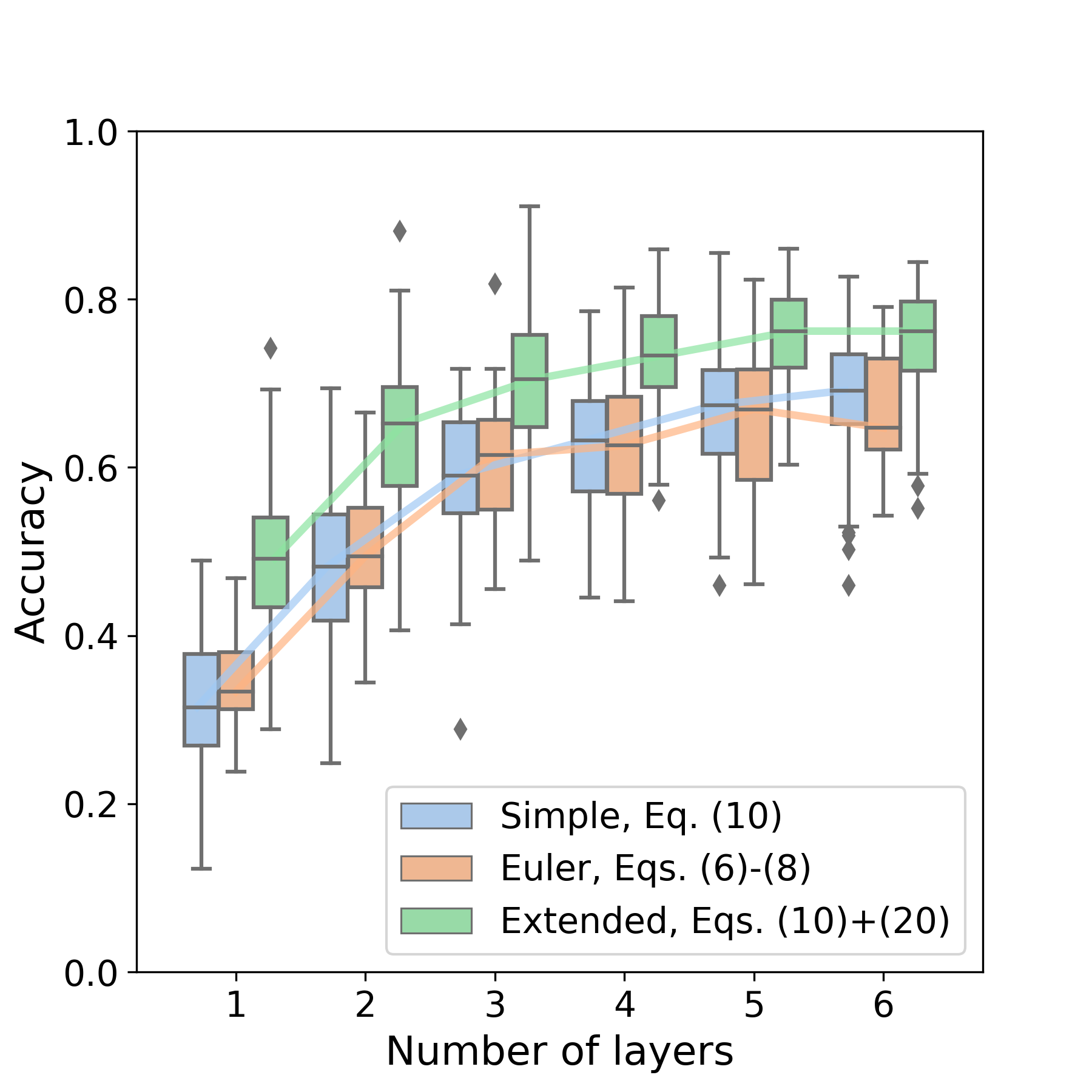}
     \includegraphics[width=0.23\textwidth]{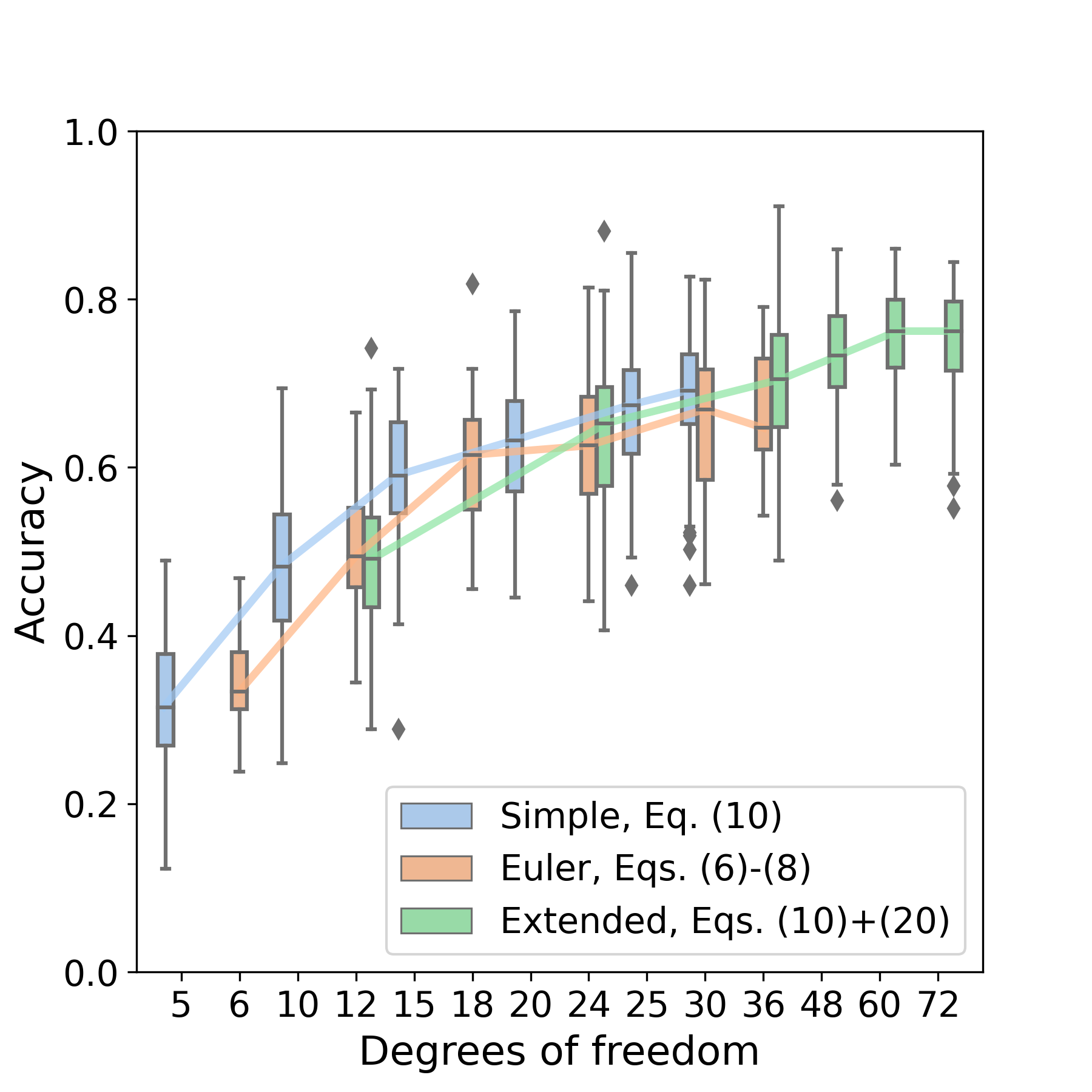}
       \caption{
       Comparison of results of qudit data re-uploading circuits of different structures on the $d=6$ class MNIST data set with input dimension $D=2$. 
       The left panels shows the results as function of the number of layers, while the right panel shows it as function of the degrees of freedom, i.e.\ the number of tunable parameters in the circuit (notice: the $x$-axis is not to scale in the right plot)
       }\label{fig:extOps}
\end{figure}

The results of the three types of circuit structures on the six-class reduced MNIST data set are shown in Fig.~\ref{fig:extOps}.
The left panel shows the accuracies as a function of the number of circuit layers. First, one can observe that there is no significant difference between the simplified structure of Eq.~\eqref{eq:simleCirc}  and circuit structure of Eqs.~\eqref{eq:DRUL_euler1}-\eqref{eq:DRUL_euler2}. On the other hand, the performance of the circuits with the extended set of operators is significantly better. 
At first glance this seems to support the hypothesis that adding more operators to the set generators does in fact enhance the trainability of qudit quantum circuits.
However, adding more operators also introduces more free parameters to the quantum circuit for a given number of layers, which allows for more flexibility in the trainable circuit.  
When comparing the accuracies as a function of number of free parameters, as shown in the right panel of Fig.~\ref{fig:extOps}, one can observe that the performance is statistically the same for all circuit structures and number of operators.
Therefore, we conclude that it is not the number of operators which is determining the trainability, but rather the number of trainable parameters.  

This reveals a tradeoff between the complexity of each layer, i.e.\ the number of operators and free parameters in each layer, and the total number of layers which are necessary to achieve a certain accuracy. 
Including more operators in each layer allows to achieve good  performance with fewer number of layers. On quantum hardware, where gate errors play an important role, the freedom to choose the elemental operations provides additional flexibility. And because different hardware may natively support different gate sets, this allows to choose configurations which result in running variational quantum circuits with less errors. 
For qubits, this is not possible since the three Pauli matrices form a basis and no additional operators can be constructed.

\subsection{Re-training on IBM hardware}\label{sec:ibmq} 

\begin{figure}
    \includegraphics[width=0.95\linewidth]{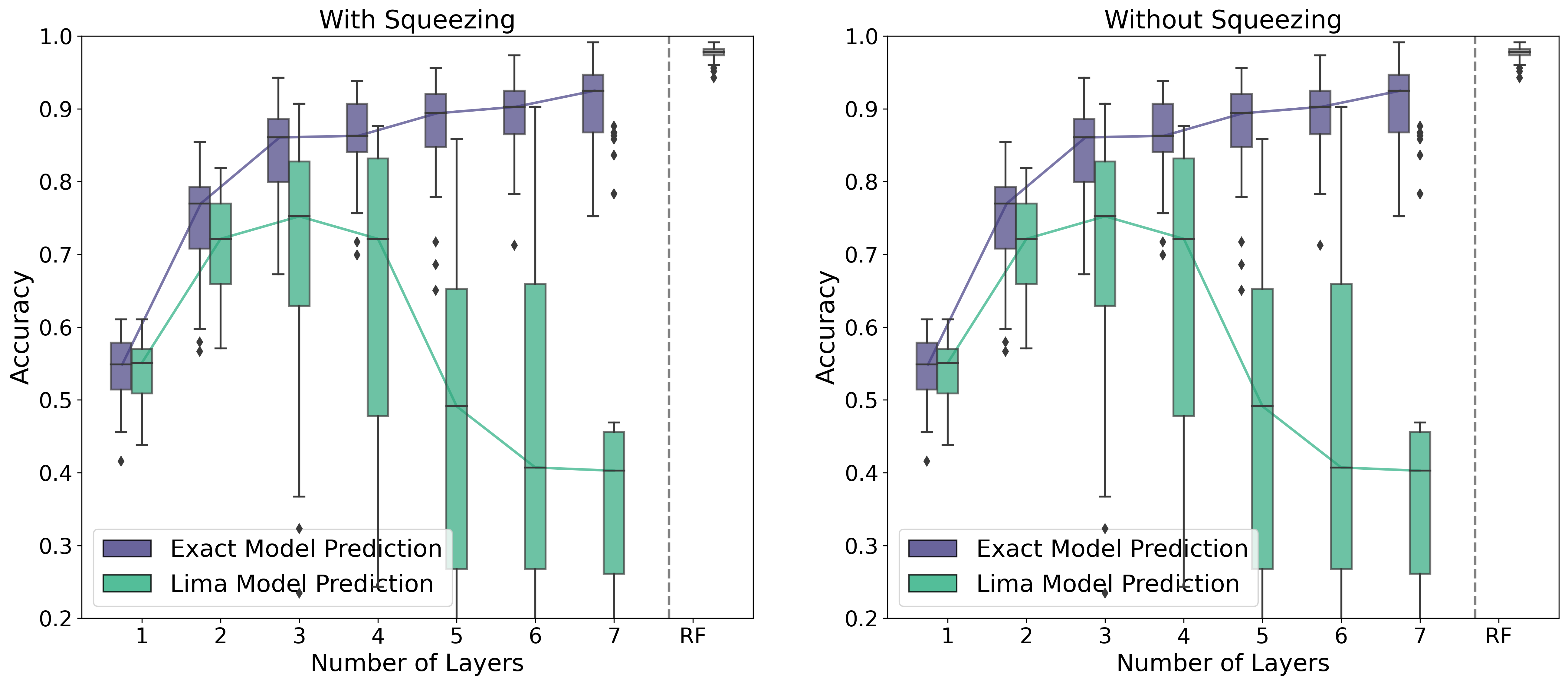}
       \caption{
       Results for the classification of the first five digits of the MNIST data set with input dimension  $D=5$ with (left) and without (right) squeezing as a function of the layers in the quantum circuit operating with a $d=5$ qudit. The numerical exact simulations are shown in blue, whereas the results from \texttt{ibmq\_lima} simulations are in green.
        The performance of a random forest (RF) model is displayed to the right in each plot
       }\label{fig:MNIST}
\end{figure}

Finally, we compare the classification performance of the model of Eqs.~\eqref{eq:DRUL_euler1}-\eqref{eq:DRUL_euler3} with and without the squeezing operation between numerically exact results and the case where the re-training and evaluation is done on actual \texttt{ibmq\_lima} hardware. For the qubit implementation of the qudit circuits we use the Dicke-state encoding described in Sec.~\ref{sec:experimental_setup}.
The left panel of Fig.~\ref{fig:MNIST} shows the accuracy as a function of the quantum circuit layers for the first five digits, i.e.\ qudit dimension $d=5$,  and with input dimension $D=5$. 
The numerically exact simulations show the same trend as previously described for two input dimensions and approach the values of the traditional RF classifier with an increasing circuit depth. However, when re-trained and evaluated on actual hardware the performance is only comparable for the first two layers, after that it saturates and then decreases strongly. 
This is due to the noise and infidelities  in the actual hardware realization of the entangling gates used to implement the squeezing operations.  
Removing the squeezing operations from the circuit reveals this explicitly, since the accuracies of both approaches are then comparable as shown in the right panel of Fig.~\ref{fig:MNIST}. 

This is an explicit example of a situation where is more efficient to increase the local Hilbert space and work with qudits, instead of increasing the number of qubits.

\section{Discussion}

In this work, we demonstrated that multi-level qudit systems are well-suited to be applied to multi-class classification problems, as each qudit basis state can naturally encode one class of the data. We implemented data re-uploading quantum circuits, where we used the angular momentum and the squeezing operators to build a universal gate set. We illustrated the capabilities of the qudit-based approach on regression and classification benchmarks. Owing to the ability to learn highly non-linear classification boundaries, the models were able to successfully learn on various data sets and achieve performances comparable to standard classical machine learning models. 
Interestingly, the achievable performance was strongly dependent on the qudit states representing the class labels and their relation to the structure of the labels in the data set. 
This intrinsic bias due to the label alignment can boost the performance of qudit circuits  substantially, when the data set is structured accordingly, which might be beneficial for certain types of application problems.

We also studied the influence of the choice of the elementary operations and the layer structure in qudit quantum circuits. 
There, we found, that the structure, i.e.\ the particular sequence and types of rotations, does not appear to have significant influence on the performance, as long as the number of degrees of freedom  was accounted for.
This reveals a trade-off between the number of elementary operators in the gate set and the number of layers to achieve the same performance. In some situations it might be advantageous to utilize the minimum set of operators and to employ more layers, while in others, more operators and less layers might be the better choice. 
This is especially interesting on quantum hardware where different gates typically have different error rates, and the freedom to choose between several configurations may allow to minimize the influence of errors on the resulting performance.

However, our data encoding layers always employed only $L_z$ and $L_x$ operators and thus implemented a ladder structure. 
This is linked to the observed intrinsic bias, and selecting different operators for the data encoding might mitigate this bias. However, we did not investigate such extensions and leave them for future work.

It should be noted that we did not employ any of the more sophisticated techniques to improve the performance of  single-qudit data re-uploading circuits, for example, employing different classical optimizers~\citep{deller_quantum_2023,lavrijsen_classical_2020}, fine-tuning hyperparameter settings~\citep{pascal_hyperparameter_2022}, and finding better parameter initializations~\citep{sack_quantum_2021,grant_initialization_2019,egger_warm-starting_2021}. All techniques have been shown to be very beneficial in related contexts and can be used in the future to improve the current approach. 

A necessary next step is to investigate the performance of multi-qudit circuits, since here we only investigated single-qudit circuits.
Interesting research questions include how the intrinsic bias of a single qudit is influences a multi-qudit circuit and its learning performance, and what the role of the set of elementary operators is in these cases. Crucially, the effect of multi-qudit entangling gates needs to be elucidated.     

In summary, our results and discussion support the conclusion that qudit systems offer a promising alternative quantum computing architecture. 
There are several differences to qubit-based systems which could potentially be leveraged to eventually provide practical benefits for quantum algorithms and quantum machine learning tasks in particular.

\bmhead{Acknowledgments}
NLW acknowledges funding from the Honda Research Institute Europe GmbH  for attending the conference  on Quantum Techniques in Machine Learning (QTML) 2022. NLW would like to thank the Van der Waals-Zeeman Institute in Amsterdam for their extended hospitality.   
SS acknowledges funding by the European Union under Horizon Europe Programme -- Grant Agreement 101080086 -- NeQST. 
Views and opinions expressed are however those of the author(s) only and do not necessarily reflect those of the European Union or European Climate, Infrastructure and Environment Executive Agency (CINEA). Neither the European Union nor the granting authority can be held responsible for them.

\section*{Declarations}
\bmhead{Conflict of interest} The authors declare no competing interests.

\bmhead{Authors' contributions}
FJ and NLW conceived the original idea on the simulations, NLW, FJ and SS conduced the numerical simulations, NLW, MSR and SS analyzed the results data. All authors jointly wrote and reviewed the manuscript.

\bmhead{Availability of data and materials}
The datasets and numerical details necessary to replicate this work are available from the corresponding author upon reasonable request.

%

\end{document}